\documentclass[12pt]{elsarticle}
\usepackage[top=2.5cm,bottom=2.5cm,left=2cm,right=1.5cm]{geometry}

\usepackage{graphicx}
\usepackage{blindtext}
\usepackage [autostyle, english = american]{csquotes}
\MakeOuterQuote{"}
\usepackage{amsmath}
\usepackage{amsfonts}
\usepackage{stfloats}
\usepackage{graphicx}
\usepackage[colorlinks]{hyperref}
\usepackage{breqn}
\graphicspath{ {./} }
\usepackage{hyperref}
\usepackage{float}
\usepackage{stfloats}
\usepackage{placeins}
\usepackage{amssymb}

\makeatletter
\def\ps@pprintTitle{%
	\let\@oddhead\@empty
	\let\@evenhead\@empty
	\let\@oddfoot\@empty
	\let\@evenfoot\@oddfoot
}
\makeatother

\begin{document}

\begin{frontmatter}



\title{Stochastic differential theory of cricket}


\author{Santosh Kumar Radha}

\address{Department of physics\\ *Case Western Reserve University}
\ead{srr70@case.edu}
\ead[https://santoshkumarradha.me]{www.santoshkumarradha.me}
\begin{abstract}

A new formalism for analyzing the progression of cricket game using Stochastic differential equation (SDE) is introduced. This theory enables a quantitative way of representing every team using three key variables which have physical meaning associated with them. This is in contrast with the traditional system of rating/ranking teams based on combination of different statical cumulants.  Further more, using this formalism, a new method to calculate the winning probability as a progression of number of balls is given. 
\end{abstract}


\end{frontmatter}


\section{Introduction}

Sports, as a social entertainer exists because of the unpredictable nature of its outcome. More recently, the world cup final cricket match between England and New Zealand serves as a prime example of this unpredictability.  Even when one team is heavily advantageous, there is a chance that the other team will win and that likelihood varies by the particular sport as well as the teams involved. There have been many studies showing that unpredictability is an unavoidable fact of sports \cite{martin}, despite which there has been numerous attempts at predicting the outcome of sports \cite{shuo}. Though different, there has been considerable efforts directed towards predicting the future of other fields including financial markets \cite{wei}, arts and entertainment award events \cite{domini}, and politics \cite{andran}.

There has been a wide range of statistical analysis performed on sports, primarily baseball and football\cite{park,iv,stokes} but very few have been applied to understand cricket. Complex system analysis\cite{muk}, machine learning models\cite{passi,ml1}  and various statical data analysis\cite{stat1,stat2,stat3,stat4,stat5} have been used in previous studies to describe and analyze cricket. In this paper, we have come up with statistical stochastic models that describe cricket. The advantage of using DEs to model events is that, it leads to making physically realistic assumptions on the sport which leads to having variables in SDEs that have describable meaning attached to it. 

We introduce the concept of using SDE for the game of cricket using a very rudimentary model in \autoref{sec:model1}. This model has a very basic analytic form and introduces the ideas behind this paper. Since SDE models the sport at its fundamental level, one can then use these models as basis for adding more terms that take into account the complexities of the game. As a example, in \autoref{sec:model2} we proceed to add slight modifications to the same model to describe in better the concept \textit{"wickets in hand"}. Example uses of these models are shown for a particular historic cricket game between \textit{India-Srilanka}. Finally a more sophisticated  model that describes the exact dynamics between the two teams is developed and shown in \autoref{sec:model3}.

\section{Formalism}

\subsection{Model 1} \label{sec:model1}
\begin{figure}[H]
	\centering
	\includegraphics[width=.6\textwidth]{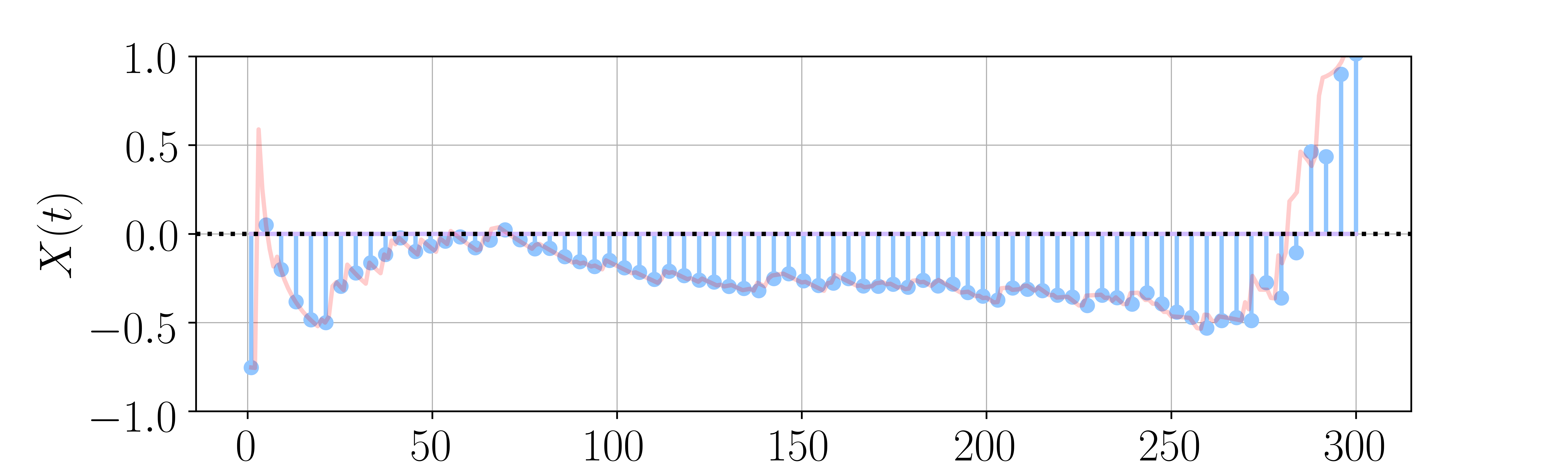}
	\caption{typical $X(t)$ from a cricket match. Here it is shown for an ODI game and hence $T=300$. Blue points represent every $4^{th}$ ball and red line, the entire game.}
	\label{fig:Xt}
\end{figure}

Central Idea behind this entire theory relies on using the variable $X$ defined as the difference between Required run rate per ball $(RR)$ and Net run rate per ball $(NR)$\footnote{RR  (Required run rate) is defined as \textit{(runs scored in first innings - runs till current ball)/No. of balls left in this innings } NR (Net run rate per ball) is defined as the  \textit{current score in second innings/number of balls played}} given by
\begin{align}
X(t)&=NR(t)-RR(t)  \label{eq:1} \\ t&\in [0,T] \notag
\end{align}
Here, $T$ refers to the total number of balls in the game. Since we are using $RR$ to define the variable, the model can be used only after the first innings of the game. This variable $X(t)$ is an quantity that is typically in between  $\pm3$ and more negative the number is, the more the team batting in the second innings is loosing. \autoref{fig:Xt} shows a typical plot $X(t)$ for an ODI game taken from a match between India and Srilanka. 

\begin{figure}[H]
	\centering
	\includegraphics[width=.5\textwidth]{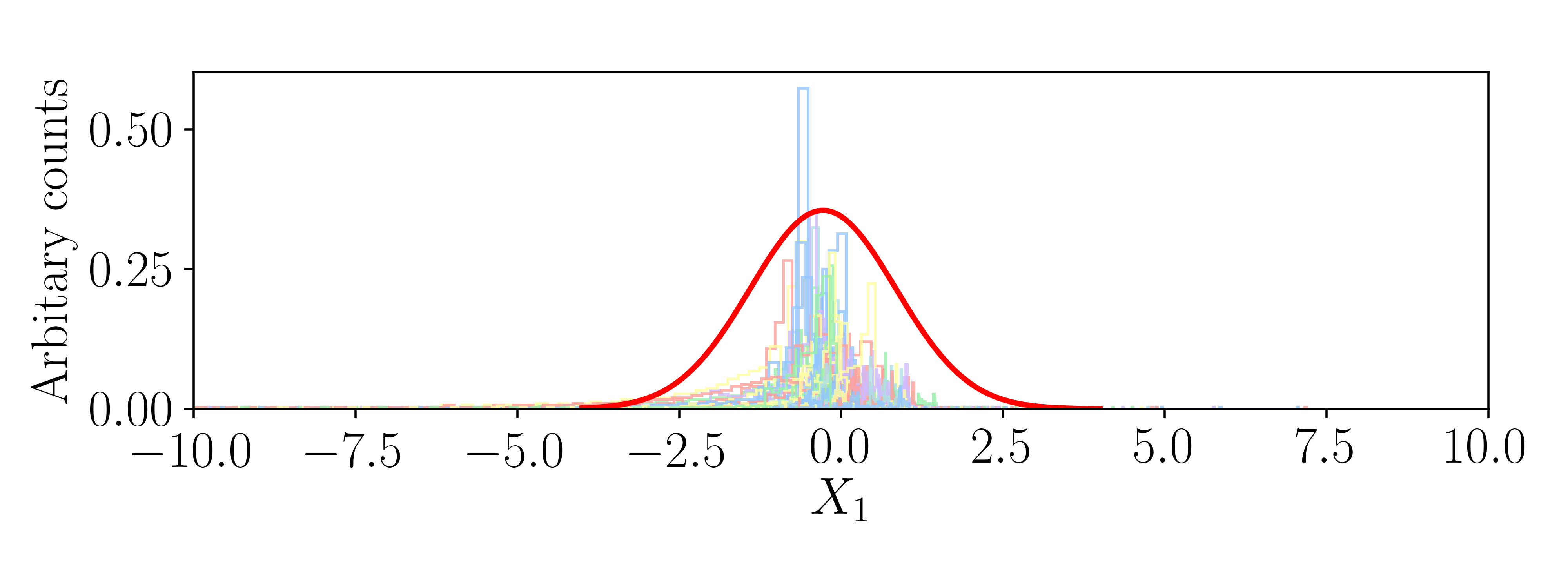}
	\caption{Actual distribution of team England of their last ball $X_1$ variable for ODI's from year $2005-2017$. This ends up with $\mu$= -0.2 , $\sigma$ = 1.12 }
	\label{fig:model-1-mean}
\end{figure}

One can write the distribution of the lead of one team over an other team ($X(t)$) at any point in the game as a weinner process. Let $(W_t)_{t \geq 0+}$ denotes a standard Brownian motion satisfying
\begin{itemize}
	\item $W_0=0$
	\item With probability 1, the function $t \rightarrow W_t$ is continuous in t
	\item The process  $(W_t)_{t\geq0+}$ has stationary, independent increments
	\item The increment $W_{t+s}-W_s=N(0,t)$ 
\end{itemize}
then the equation
\begin{align}
X(t)&=\mu t + \sigma W_t \approx N(\mu t,\sigma^2 t) \label{eq:2} 
\end{align} 

can be used to describe the distribution of $X(t)$ with $\mu$ and $\sigma$ generalizing the winner process giving raise to mean and variance of the resulting normal distribution. \autoref{fig:model-1-mean} shows actual data fitted to this kind of process. Though the fit does not seem to be perfect, this can be used to illustrate the idea behind this formalism (Better model is developed later in the paper ). $\mu$ and $\sigma$, in stochastic terms represent the drift and volatility in the process. This immediately shows the huge advantage of modeling the sport using this type of process. $\mu$'s of each team would indicate quantitatively, the advantage(disadvantage) the team 1(2) has over team 2(1). $\sigma$ variable indicates the  degree of unpredictability the team has while playing, in-fact often times, teams become so unpredictable that they win(loose) a loosing(winning) game.

\autoref{eq:2} is nothing but a solution to the stochastic differential equation 
\begin{align}
dX_t = \mu dt + \sigma dW_t\label{eq:3} 
\end{align} 
The reason for choosing the above form of \autoref{eq:2} is the fact that this results in simple and elegant analytical solutions of various derived quantities while giving physically relevant meaning to the used parameters and variables.

\begin{figure}[htbp]
	\centering
	\includegraphics[width=\textwidth]{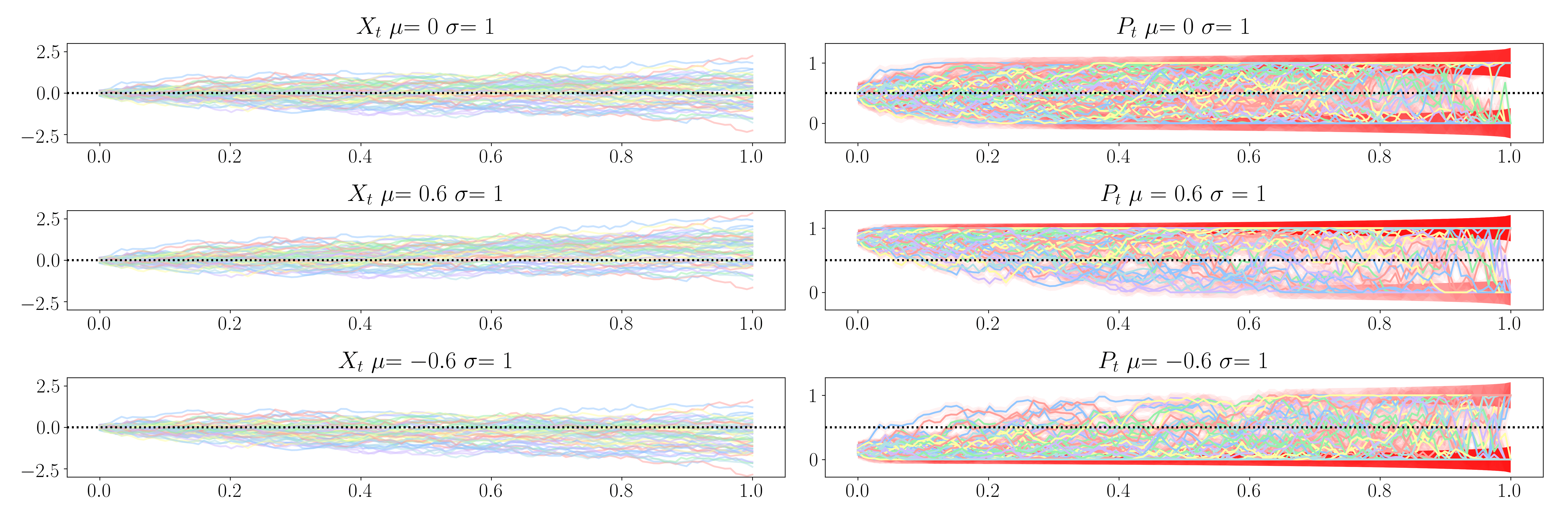}
	\caption{Simulated trajectories of \autoref{eq:3} and their corresponding  $\mathbb{P}(X(1)|X(t_1)=X_{t_1})>0$ for various values of $\mu$. Red shaded region shows the variance of each individual trajectories overlaid on each other.  }
	\label{fig:model-1}
\end{figure}

One of the important quantity we are interested in is $P_T := \mathbb{P}(X(T)>0)$. $P_T$ is the probability of the given path to reach a point above $0$. This would mean that the $NR$ is greater than $RR$ at the end of the game, thus the probability of team 2 winning. Without loss of generality, we can re-scale the time to $[0,1]$ and notice that $X(1)$ is nothing but $N(\mu,\sigma)$ and thus we get
\begin{align}
P_1 := \mathbb{P}(X(1)>0)=\frac{1}{2}\left[1+erf\left(\frac{1-\mu}{\sigma \sqrt{2}}\right)\right] \label{eq:4} 
\end{align} 

One can use this to calculate the more interesting conditional probability $\mathbb{P}((X(1) >0|X(t_1)=\alpha)$. One can easily realize that 
\begin{align}
X(1)=\mu (1-t_1) +\sigma (W_{1-{t_1}})+\alpha\label{eq:5} 
\end{align} 
\autoref{eq:5} uses the fact that $W_1-W_{1-t_1}=W_{1-{t_1}}$. Hence $P(X(1)|X(t_1)=\alpha)$ is nothing but $N(\mu (1-t_1)+\alpha,\sigma^2(1-t_1))$. Hence $P(X(1)|X(t_1)=\alpha)>0$ and any $t_1<1$ is given by
\begin{align}
\frac{1}{2}\left[1+erf\left(\frac{1-\mu (1-t_1)-\alpha}{\sigma \sqrt{2(1-t_1)}}\right)\right] \label{eq:6} 
\end{align} 

After few steps of calculations one can get a closed form solution for the variance of the above probability as 

\begin{dmath}
		\frac{1}{2} \left(erf\left(\frac{{\mu} (1-t)+{\mu}}{\sqrt{{\sigma}^2 (1-t)+{\sigma}^2}}\right)+1\right)-\left[\frac{1}{4} \left(erf\left(\frac{{\mu} (1-t)+{\mu}}{\sqrt{{\sigma}^2 (1-t)+{\sigma}^2}}\right)+1\right)^2+2 \mathcal{T} \left(\frac{(1-t) {\mu}+{\mu}}{\sqrt{(1-t) {\sigma}^2+{\sigma}^2}}-\frac{{\sigma} \sqrt{1-t}}{\sqrt{(1-t) {\sigma}^2+2 {\sigma}^2}}\right)\right]  \label{eq:7}
\end{dmath}

where  $\mathcal{T}$ is the Owen's T function. It is interesting to note that the variance is independent of the value at time $t$.

\autoref{fig:model-1} shows the plot of trajectories simulated using \autoref{eq:3} and their corresponding values of probability of the trajectory's value being $>0$ given the trajectory's value at time $t$ is $X_t$. Top panel shows the trajectories for $\mu=0$, as one can see on an \textit{"average"} the final path is equally split between being positive and negative and thus winning and loosing. This would mean that the team has no advantage whatsoever compared to the other team it is playing and it is reflected clearly in the probability graph too. When plotted for high enough simulated trajectories, number of trajectories reaching probabilities 0 and 1 are the same. This symmetry in probability and trajectory can also be seen from \autoref{eq:3} by setting $\mu$ to 0. Contrastingly , bottom two panels show for positive and negative value of $\mu$ and one can see the clear anisotropy as expected in the probability graphs with the final $X(1)$ on a average being around $\mu$. $\sigma$ describes the ability for how \textit{"tough"} the competition is going to be. For instance $\mu$ of 0.4 and $\sigma$ of 3 would imply that team 2, even though being advantageous than team 1, has a high variance  of either loosing or winning compared to the team having $\sigma=1$ with same $\mu$.  

\autoref{fig:model-1-mean} shows the distribution of $X_1$ for team England for the period of $2005-2017$ for all their ODI games and fitting the normal curve as shown in previous paragraphs, one can see that England has $\mu$= -0.2 and $\sigma$ = 1.12. This might seem a bit odd as this shows England, as a team, for the given period is at a disadvantage. This is so because we have fitted the normal curve to all of England's games but rather in actual scenario, one has to fit the data of team 1 against team 2 as relative difference between $\mu$ and $\sigma$ makes more sense. A test case of England against Pakistan gives a $\mu$ of 0.17 which shows that when England and Pakistan play a game, England has an advantage, at least according to our model.

\autoref{fig:results} shows  the calculated probability for an actual game between India and Sri lanka as an example. Top panel shows the progression of $X(t)$ and the second from top shows $P(t)$ using the model written above. Naively one would expect that a negative value of $X(t)$ would indicate a disadvantage for the chasing team and a positive value, advantage. This plot shows the root of the concept proposed in this paper. We have $X$, which is a measurable function $X:\Omega \rightarrow S$ where $\Omega \in [0,1]$ and $E\in [-\infty,\infty]$.  We then use the fact we have a probability measure on $(\Omega,\mathcal{F})$ and then define 

\begin{align}
P(X \in E)=P({t \in \Omega | X( t ) \in E}) \label{eq:8} 
\end{align}
we then go to a new subspace which is made up of $P(t)$ (given in \autoref{eq:6}) as our random variable and use $P(X(1)|X(t_1)=X_{t_1})>0$ as our measure. Thus we have created a mapping of measure function from $X \rightarrow P$ where the measure space has moved from $E \in  [-\infty,\infty]$ to $E^\prime  \in [0,1]$. This is clearly reflected in \autoref{fig:results} from comparing First and second panel (from top) as the shape of the curves remain essentially the same. Second panel shows the calculated probability with red and green shaded areas indicating a probability lesser or greater than $0.5$.  Probabilities are calculated using $\mu$ and $\sigma$ fitted for all games of India and Sri lanka as explained above.  This graph answers this question - Given that India is playing against Sri lanka, what is the probability of India winning the game given that the difference between Net run rate and Required run rate is $X_t$ at ball $t$.

\subsection{Model 2} \label{sec:model2}
\begin{figure}[htbp]
	\centering
	\includegraphics[width=\textwidth]{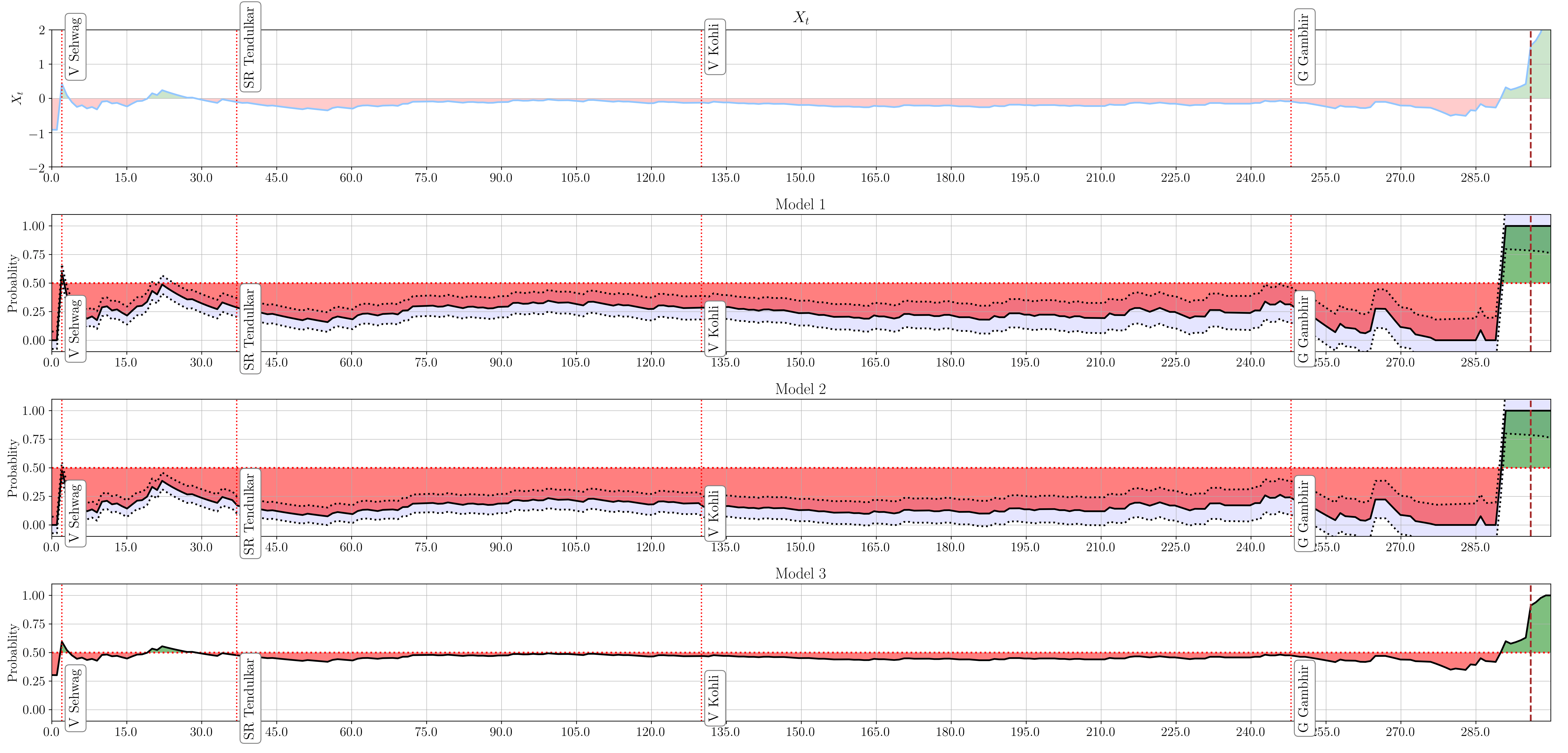}
	\caption{$X_t$ and $P_t$ calculated using the models shown in the paper for a real game of India against Sri lanka (\textit{ICC Cricket World Cup at Mumbai, Apr 2 2011}). Each vertical red line indicate the player getting out. Dotted area shows the variance and the confidence level of the calculated probabilities. Final red line indicate the game finishing before 300 balls. }
	\label{fig:results}
\end{figure}

Vertical lines in \autoref*{fig:results} refer to the balls at which different wickets have fallen. This leads to the next modification that one can make to the model. Till now the model has been represented by the \textit{Macro} effects of the game like the past wins and past data, but one does know that loosing wickets in middle of the game perturbs  the system and pushes it away from equilibrium. Ideally this would lead to \autoref{eq:3} changing to 
\begin{align}
dX_t = \mu(t) dt + \sigma(t) dW_t\label{eq:9} 
\end{align}

One can model $\mu(t)$ by the following method, 

\begin{align}
\mu(t)&= \mu -|\bar{\mu}|f(\bar{w},w_t) \label{eq:10}  \\
f(x,\lambda)&=1-\sum_{i=0}^{x}\frac{e^{-\lambda}\lambda^i}{i!} \notag
\end{align} 

$\bar{\mu}$ is called the \textit{disadvantage factor}  which is a universal constant for the game,$\bar{w}$ the average number of wickets lost by the team in the past games and $w_t$ the number of wickets remaining at time $t$. $f$ is nothing but the survival function of the Poisson distribution. The assumption we have made here is that the fall of wickets follows a Poisson distribution. 
\begin{figure}[H]
	\centering
	\includegraphics[width=.7\textwidth]{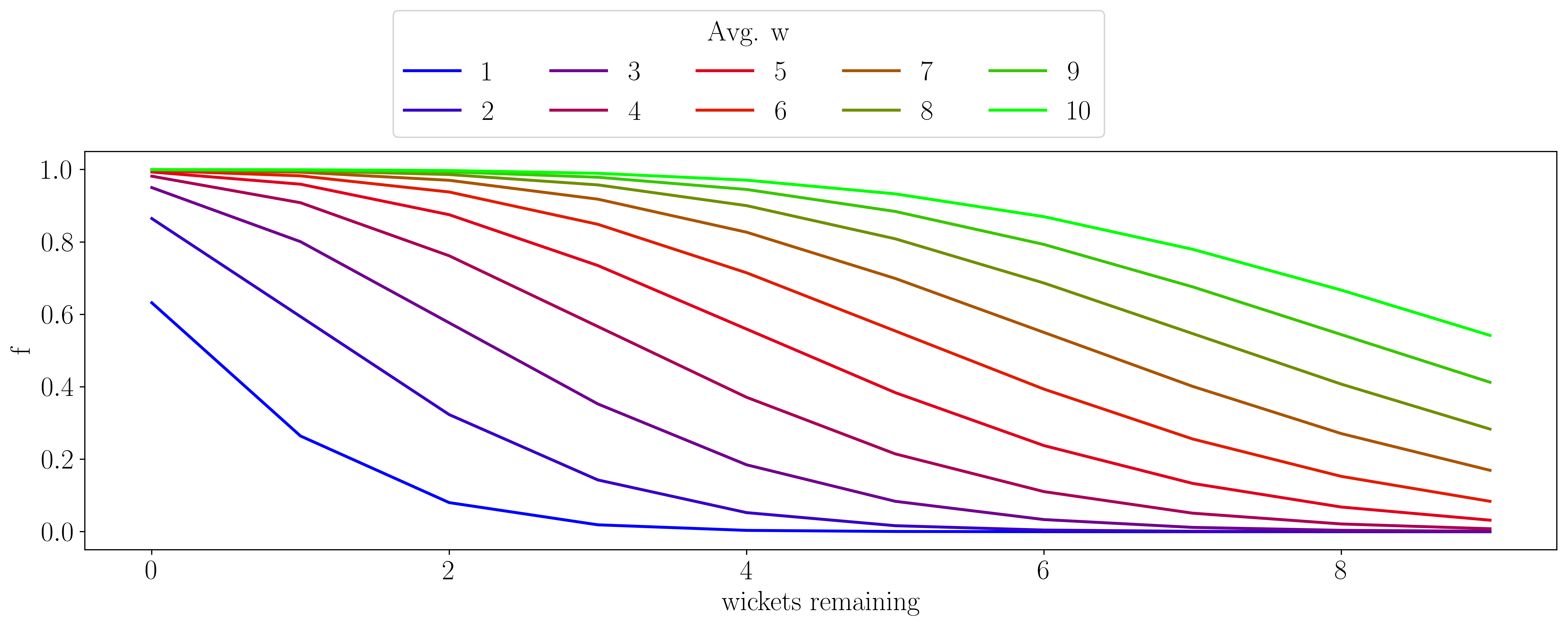}
	\caption{$f(\bar{w},w_t)$ from \autoref{eq:10} for various mean wickets lost in a game. }
	\label{fig:wickets-model-2}
\end{figure}

\begin{figure}[H]
	\centering
	\includegraphics[width=.7\textwidth]{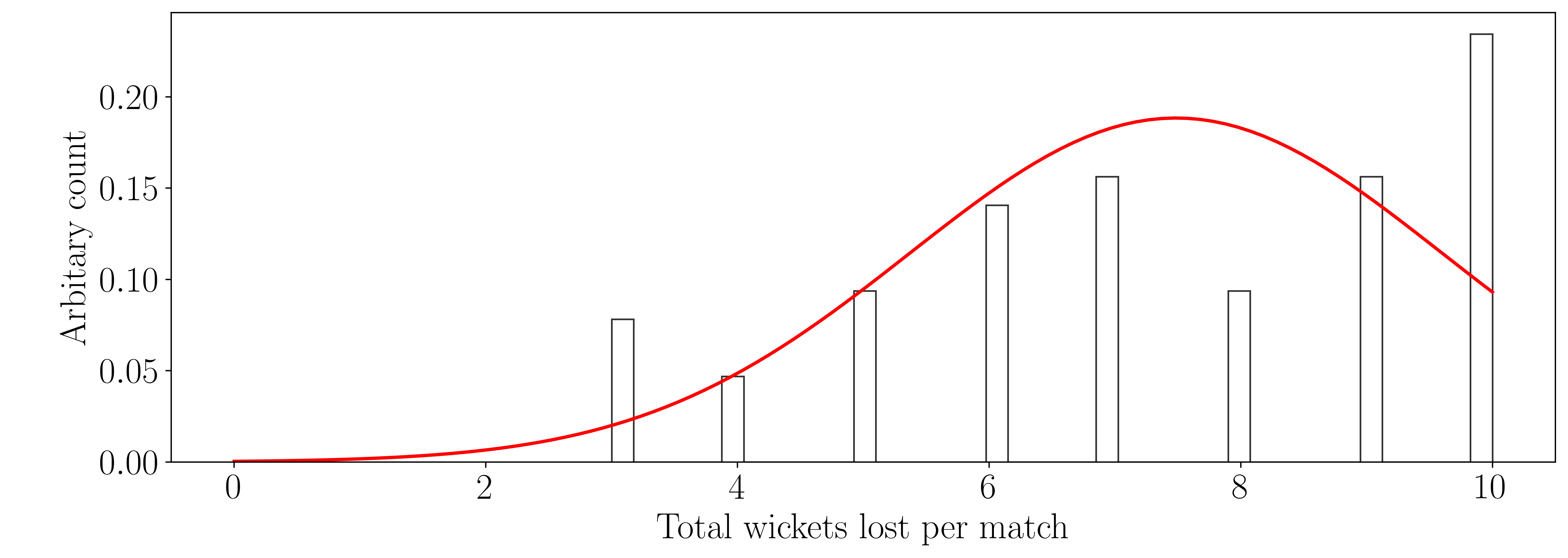}
	\caption{Distribution of actual wickets lost by team India from the period  $2005-2017$. Red curve shows the normal fit. }
	\label{fig:wickets-india-model-2}
\end{figure}

\autoref{fig:wickets-model-2} shows the function $f$ for various average number of wickets lost. One can look at the extreme cases to understand the behavior of the function. For  average number of wickets lost ($\bar{w}$) $=10$ the team starts with a high disadvantage even when no wicket is lost. But a team with $\bar{w}$ $=1$ has a very low disadvantage contribution even when they have lost $9$ wickets. Thus for a given team  $\bar{w}$ determines the perturbation caused by the loss of a wicket when the remaining wickets are $w_t$ at time $t$ given by  $f(\bar{w},w_t)$. After perturbing $\mu$ we make another simplification and consider a \textit{Born Oppenheimer approximation} to \autoref{eq:8} and directly calculate the $P_t$ using previous equations. \autoref{fig:wickets-india-model-2} shows the distribution of wickets lost for each game for team India from $2005-2017$. India losses on an average 7.4 wickets a game with a variance of 2.11. This would correspond to a curve in-between 7 and 8 in \autoref{fig:wickets-model-2}. 

\autoref{fig:results} 's third panel from top (model 2) shows the probability calculated using this new scheme. One can see immediately in the first ball that the probability of winning is increased compared to model 1 as we know that India only loses an average of 7 wickets every game  and thus without loss of any wicket, we have a higher chance of winning. Second feature to notice is the sudden response to the applied perturbation after each wicket, sharp decrease in the probability followed by relaxation.

\subsection{Model 3} \label{sec:model3}

Previous models relaid on using \autoref{eq:9} as the basis of evolution of $X_t$, though this gave us good insights on how to use SDE's to model cricket, this rudimentary equation lacks detail. Previous models fail to capture one important fact of the game where each team tries to play better as they become more advantageous along the game. For instance a team winning in mid game, would have a boosted morale to win the game than the opposition who is in the loosing spot . This translates into having the following SDE for $X_t$
\begin{align}
dX_t=x_0x_1-x_0X_tdt+\sigma dW_t \label{eq:11} 
\end{align}

This is almost a Ornstein–Uhlenbeck\cite{ou} process with slight modifications and one can get the solution of it with few simple steps as 
\begin{align}
X_t=X_{t-1}(e^{{-x_0t}})+x_1\left(1-e^{{-x_0t}}\right)+\sigma e^{{-x_0t}}\int _{0}^{t}e^{{x_0s}}\,dW_{s} \label{eq:12} 
\end{align}

From \autoref{eq:12} one can calculate the expectation value and variance of the same and easily arrive at 
\begin{align}
{E}[X_{t}]&=X_{t-1}e^{{-x_0t}}+x_1(1-e^{{-x_0t}}) \label{eq:13} \\
{Var}[X_{t}]&={\frac {\sigma ^{2}}{2x_0}}(1-e^{{-2x_0t}})\label{eq:14}
\end{align}
Similar to previous models, we can get the probability of reaching a positive number at $t=1$ given that at time $t$ the value is $\alpha$,  $\mathbb{P}(X(1)>0|X(t)=\alpha)$ as
\begin{dmath}
	\mathbb{P}(X(1)>0|X(t)=\alpha) = \\  \frac{1} {2}\text{ erf}\left(\frac{\alpha \sqrt{2}  e^{x_1 \left(1-e^{x_0 (t-1)}\right)+(t-1) x_0}}{\sigma \sqrt{\frac{1-e^{2 (t-1) x_0}}{x_0}}}\right)+\frac{1}{2} \label{eq:15}
\end{dmath}
\autoref{eq:15} is the final probability of winning the game using this model. Note that we derived the probability after rescaling $t$ variable in  \autoref{eq:11}  from $[0,\infty] \rightarrow [0,1]$. One striking feature of this process is that (from \autoref{eq:13})
\begin{align}\
\lim_{{t\to \infty(1)}}{\mathrm {E}}[X_{t}]&=x_1 \label{eq:16}\\
\lim_{{t\to \infty(1)}}{\mathrm {Var}}[X_{t}]&={\frac {\sigma ^{2}}{2x_0}} \label{eq:17}
\end{align}

\begin{figure}[!htbp]
	\centering
	\includegraphics[width=\textwidth]{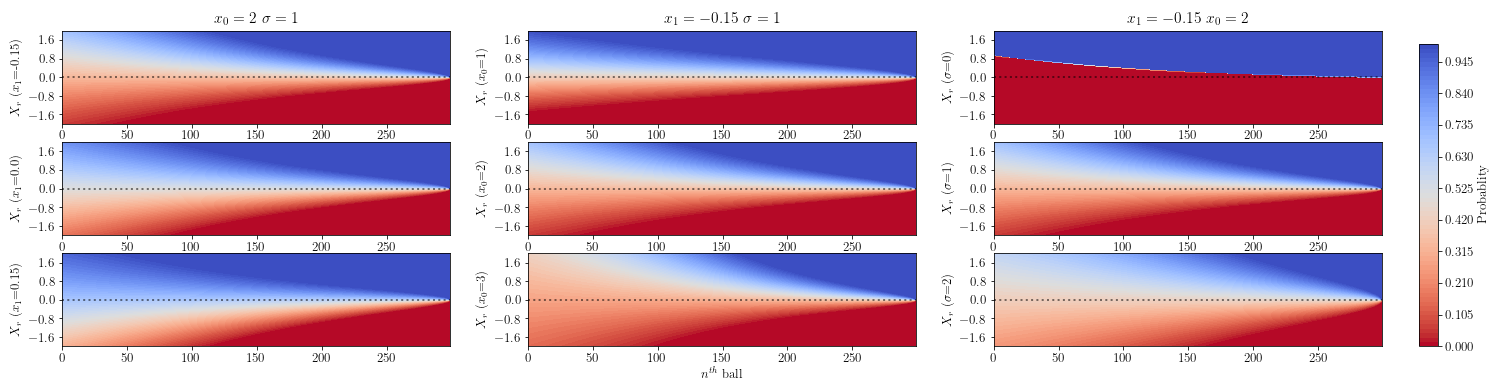}
	\caption{ Color map of $P_t$ calculated for various values of $x_0,x_1,\sigma$ for \autoref{eq:11} with $X_t$ being constant across all balls. Dotted lines indicate $P_t$ when $X_t=0$ for all $t$.  }
	\label{fig:model-3}
\end{figure}

This clearly elucidates the physical meaning of $x_0,x_1$ and $\sigma$. \autoref{fig:model-3} shows the calculated probability for various values of $x_0,x_1$ and $\sigma$ for a constant value of $X_t$ thorough out the game (for \textit{ex.} $X_t$ is 0.4 from ball 1 to ball 300). 

\begin{figure}[h]
	\centering
	\includegraphics[width=.6\textwidth]{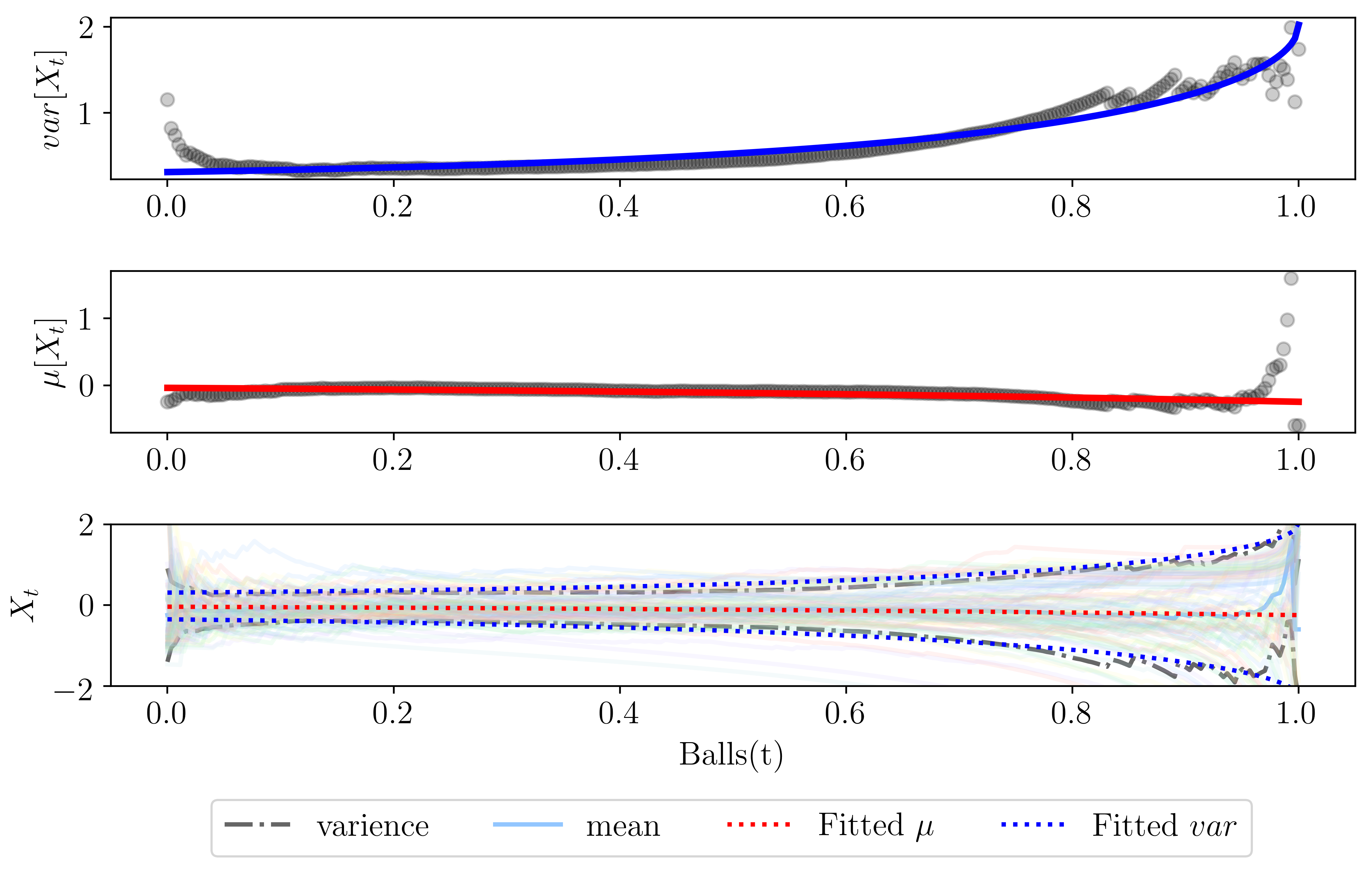}
	\caption{ Fit for mean and variance for team India using \autoref{eq:13} and \autoref{eq:14} for the time period of $2005-2017$. Top and middle panels shows the fit for variance and mean. Lowermost panel shows the fitted curve along with actual curve with actual trajectories of all games India has played from  $2005-2017$ overlaid on top of them}
	\label{fig:fit-model-3}
\end{figure}

One can now use actual data to calculate $var[X_t]$ and $\mu[X_t]$ as a function of number of balls by including all games $\forall t \in [0,1]$ and then fit it to \autoref{eq:13} and \autoref{eq:14} to extract  $x_0,x_1$ and $\sigma$. \autoref{fig:fit-model-3} shows an example for this kind of fit for team India ($x_0$=1.18 $x_1$=0.06 $\sigma$=-2.76). Bottom panel shows the fitted curve along with mean and variance of actual data with all the data overlaid for reference. As seen fro the fit, this model seems to model the dynamics of the game perfectly than our previous one. 

Bottommost panel in \autoref{fig:results} shows the model applied to the game described above with the fitted variables. This plot shows that from start, India, though seemed like are on the loosing side just from $X_t$ (topmost panel), are actually not that bad with probability of 0.5 almost throughout the game

\begin{figure}[!tbh]
	\centering
	\includegraphics[width=.6\textwidth]{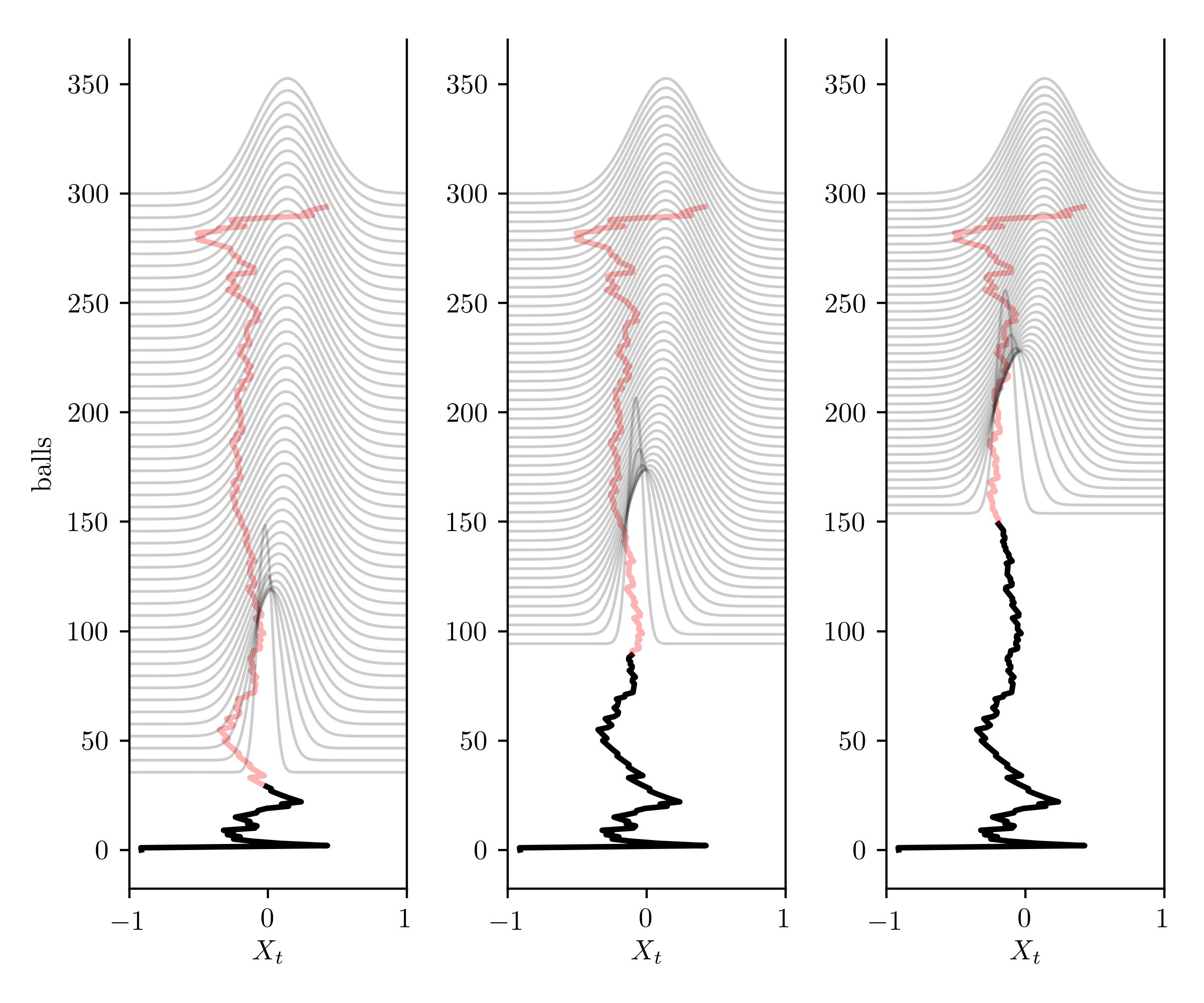}
	\caption{ Thick black and red curves shows the actual $X_t$ from a cricket game between India and Sri lanka. Black distributions are solutions to the equation of motion overlaid on top for every ball from a)$ 10^{th} $ b)$ 90^{th} $ c)$ 150^{th} $ }
	\label{fig:example-path	}
\end{figure}

As mentioned earlier, this paper boils down to mapping $X_t : \rightarrow \mathcal{P}_t$.  Model 3 developed in \autoref{sec:model3} can now be used in predicting the probability of winning the game at every ball. \autoref{fig:example-path	} shows the model in action at ball $10,90$ and $150$ for the same game mentioned above. It shows the evolution of the probability from the current known data. Black and red thick line shows the actual data and the thin black lines show the expected distribution of $X_t$ for each ball. One can clearly see the evolution of the probability starts with a $\delta$ function which slowly stars to spread out with time(ball) varying mean and variance. This essentially can be derived from Fokker–Planck representation of \autoref{eq:11}.
\section{Conclusion}

We have introduced a a new formalism for understanding the progression of cricket using underlying variables in the game. We show that it boils down to using $X_t:=RR-NR$ as the fundamental stochastic variable defining the progression of game and each team trying to either make $X_t > 0$ or $< 0$. Using this premise, three models of various complexities are developed to show the versatility of using SDEs to model the sport. One interesting application of using the models a predictive indicator by calculating probability of winning is shown. One can now use the variables defined in \ref{sec:model3} ($\vec{X}:=(x_0,x_1,\sigma$)  as a quantitative means of measuring the relative performance of each team in the sport, discussions about this would be included in a future paper.

\FloatBarrier

\section*{References} 
\bibliographystyle{elsarticle-harv}

\end{document}